\documentclass[nofootinbib,twocolumn,preprintnumbers]{revtex4}
\usepackage{amsmath}
\usepackage{amssymb}
\usepackage{epsfig}
\usepackage{graphicx}
\usepackage{inputenc}
\inputencoding{latin1}

\newcommand{\newc}{\newcommand}

\newcommand{\filt}{\mathrm{filt}}
\newcommand{\order}[1]{{\cal O}\left(#1\right)}
\newcommand{\cut}{\mathrm{cut}}

\newc{\Lumi}{{\cal L}}

\newc{\ra}{\rightarrow}
\newc{\Ra}{\Rightarrow}

\newc{\pom}  {I\hspace{-0.2em}P}

\newc{\gev}{{\rm GeV}}

\newc{\rpv}{{\not \!\! R_p}}
\newc{\rpvm}{{\not  R_p}}

\newc{\gsim}{{\stackrel{>}{\sim}}}
\newc{\lsim}{{\stackrel{<}{\sim}}}
\newc{\sleq} {\raisebox{-.6ex}{${\textstyle\stackrel{<}{\sim}}$}}
\newc{\sgeq} {\raisebox{-.6ex}{${\textstyle\stackrel{>}{\sim}}$}}

\def\3{\ss}

\newc{\ETJ}{E^{{\rm jet}}_T}

\def\ptmin{\hat{p}_T^{\rm min}}

\def\pt{p_T}

\def\kt{K_\perp~}

\def\q2{{\rm Q}^2}
\def\p2{{\rm P}^2}
\def\gev2{{\rm GeV}^{2}}
\def\F2g{F_2^\gamma}
\def\f2c{F_2^{\rm charm}}

\def\d0{D^{0}}

\def\begr{\begin{flushright}}
\def\endr{\end{flushright}}

\def\begl{\begin{flushleft}}
\def\endl{\end{flushleft}}

\def\herwig{HERWIG}
\def\pythia{{\sc Pythia}}

\def\kt{$K_\perp$}

\newcommand{\as}{\alpha_s}

\DeclareMathAlphabet{\mathsc}{OT1}{cmr}{m}{sc}

\newcommand{\mh}{m_{\mathsc{h}}}

\newcommand{\GeV}{\,\mathrm{GeV}}

\begin{document}

\title{Jet substructure as a new Higgs search channel at the LHC}

\author{Jonathan~M.~Butterworth, Adam~R.~Davison}
\affiliation{Department of Physics \& Astronomy, University College
  London.}
\author{Mathieu~Rubin, Gavin P.~Salam}
\affiliation{LPTHE; UPMC Univ.\ Paris 6; Univ.\ Denis Diderot; CNRS UMR
  7589; Paris, France.}

\begin{abstract}
  It is widely considered that, for Higgs boson searches at the Large
  Hadron Collider, $WH$ and $ZH$ production where the Higgs boson
  decays to $b\bar{b}$ are poor search channels due to large
  backgrounds.  We show that at high transverse momenta, employing
  state-of-the-art jet reconstruction and decomposition techniques,
  these processes can be recovered as promising search channels
  for the standard model Higgs boson around 120 GeV in mass.
\end{abstract}


\maketitle

A key aim of the Large Hadron Collider (LHC) at CERN is to discover
the Higgs boson,
the particle at the heart of the standard-model (SM) electroweak
symmetry breaking mechanism. 
Current electroweak fits, together with the LEP exclusion limit,
favour a light Higgs boson, i.e.\ one around 120 GeV in
mass~\cite{Grunewald:2007pm}. This mass region is particularly
challenging for the LHC experiments, and any SM
Higgs-boson discovery is expected to rely on a combination of several
search channels, including gluon fusion $\to H \rightarrow
\gamma\gamma$, vector boson fusion, and associated production with
$t\bar{t}$ pairs~\cite{atlasphystdr,cmsphystdr}.

Two significant channels that have generally been considered less
promising are those of Higgs-boson production in association with a
vector boson, $pp\to WH$, $ZH$,
followed by the dominant light Higgs boson decay, to two $b$-tagged
jets.
If there were a way to recover the $WH$ and $ZH$ channels it could
have a significant impact on Higgs boson searches at the LHC.
Furthermore these two channels also provide unique information on the
couplings of a light Higgs boson separately to $W$ and $Z$ bosons.

Reconstructing $W$ or $Z$ associated $H\to b\bar b$ production would
typically involve identifying a leptonically decaying vector boson,
plus two jets tagged as containing $b$-mesons.
Two major difficulties arise in a normal search scenario. The first is
related to detector acceptance: leptons and $b$-jets can be
effectively tagged only if they are reasonably central and of
sufficiently high transverse momentum. The relatively low mass of the
$VH$ (i.e.\ $WH$ or $ZH$) system means that in practice it can be
produced at rapidities somewhat beyond the acceptance, and it is also not
unusual for one or more of the decay products to have too small a
transverse momentum.
The second issue is the presence of large backgrounds with intrinsic
scales close to a light Higgs mass. For example, $t \bar t$ events can
produce a leptonically decaying $W$, and in each top-quark rest frame,
the $b$-quark has an energy of $\sim 65 \GeV$, a value uncomfortably
close to the $m_H/2$ that comes from a decaying light Higgs boson. If
the second $W$-boson decays along the beam direction, then such a
$t\bar t$ event can be hard to distinguish from a $WH$ signal event.

In this letter we investigate $VH$ production in a boosted regime, in
which both bosons have large transverse momenta and are back-to-back.
This region corresponds to only a small fraction of the total $VH$
cross section (about $5\%$ for $p_T > 200 \GeV$), but it has
several compensating advantages: (i) in terms of acceptance, the
larger mass of the $VH$ system causes it to be central, and the
transversely boosted kinematics of the $V$ and $H$ ensures that their
decay products will have sufficiently large transverse momenta to be
tagged; (ii) in terms of backgrounds, it is impossible for example for
an event with on-shell top-quarks to produce a high-$p_T$ $b\bar b$
system \emph{and} a compensating leptonically decaying $W$, without
there also being significant additional jet activity; (iii) the $HZ$
with $Z \to \nu \bar \nu$ channel becomes visible because of the large
missing transverse energy.

One of the keys to successfully exploiting the boosted $VH$ channels
will lie in the use of jet-finding geared to identifying the
characteristic structure of a fast-moving Higgs boson that decays to
$b$ and $\bar b$ in a common neighbourhood in angle.
We will therefore start by describing the method we adopt for this,
which builds on previous work on heavy Higgs decays
to boosted W's~\cite{Seymour:1993mx},
WW scattering
at high energies~\cite{Butterworth:2002tt} and the analysis of SUSY decay
chains~\cite{Butterworth:2007ke}.
We shall then proceed to discuss event generation, our precise cuts
and finally show our results.


\begin{figure*}[t]
  \centering
  \includegraphics[width=0.8\textwidth]{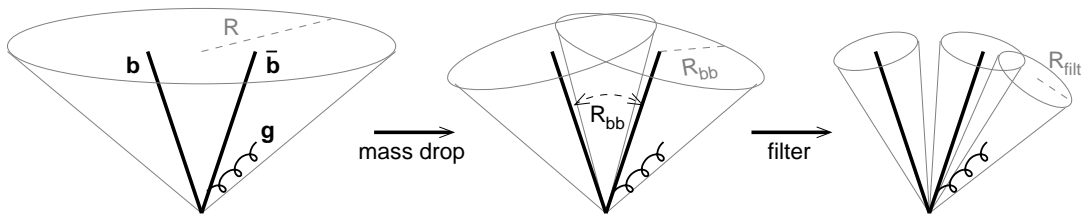}
  \caption{The three stages of our jet analysis: starting from a hard
    massive jet on angular scale $R$, one identifies the Higgs neighbourhood
    within it by undoing the clustering (effectively shrinking the jet radius) until the jet splits into
    two subjets each with a significantly lower mass; within this
    region one then further reduces the radius to $R_\filt$ and takes
    the three hardest subjets, so as to filter away UE contamination
    while retaining hard perturbative radiation from the Higgs decay
    products.}
  \label{fig:higgs-neighbour+filter}
\end{figure*}

When a fast-moving Higgs boson decays, it produces a single fat jet
containing two $b$ quarks. A successful identification strategy should
flexibly adapt to the fact that the $b\bar b$ angular separation will
vary significantly with the Higgs $p_T$ and decay orientation, roughly
\begin{equation}
  \label{eq:Rbb}
  R_{b\bar b} \simeq \frac{1}{\sqrt{z(1-z)}} \frac{\mh}{\pt}\,,\qquad
  (\pt \gg \mh)\,,
\end{equation}
where $z$, $1-z$ are the momentum fractions of the two quarks. In particular
one should capture the $b, \bar b$ and any gluons they emit, while
discarding as much contamination as possible from the underlying event
(UE), in order to maximise resolution on the jet mass. One should
also correlate the momentum structure with the directions of the two
$b$-quarks, and provide a way of placing effective cuts on the $z$
fractions, both of these aspects serving to eliminate backgrounds.

To flexibly resolve different  angular scales we use the inclusive,
longitudinally invariant Cambridge/Aachen (C/A)
algorithm~\cite{Dokshitzer:1997in,Wobisch:1998wt}: one calculates the angular distance
$\Delta R_{ij}^2 = (y_i-y_j)^2 + (\phi_i - \phi_j)^2$ between all
pairs of objects (particles) $i$ and $j$, recombines the closest pair,
updates the set of distances and repeats the procedure until all
objects are separated by a $\Delta R_{ij} > R$, where $R$ is a
parameter of the algorithm. It provides a hierarchical structure for
the clustering, like the \kt algorithm~\cite{Catani:1993hr,Ellis:1993tq},
but in angles rather than in relative transverse momenta (both are
implemented in FastJet~2.3\cite{Cacciari:2005hq}).

Given a hard jet $j$, obtained with some radius $R$, we then use the
following new iterative decomposition procedure to search for a generic
boosted heavy-particle decay. It involves two dimensionless parameters, $\mu$
and $y_\cut$:
\begin{enumerate}
\item Break the jet $j$ into two subjets by undoing its last stage of
  clustering. Label the two subjets $j_1, j_2$ such that $m_{j_1} >
  m_{j_2}$.
\item If there was a significant mass drop (MD), $m_{j_1} < \mu m_{j}$, and
  the splitting is not too asymmetric,  $y= \frac{\min(p_{tj_1}^2,
    p_{tj_2}^2)}{m_j^2} \Delta R_{j_1,j_2}^2 > y_\cut$,
  then deem $j$ to be the heavy-particle neighbourhood and exit the
  loop. Note that $y \simeq
  \min(p_{tj_1},p_{tj_2})/\max(p_{tj_1},p_{tj_2})$.\footnote{Note also
    that this $y_\cut$ is related to, but not the same as, that used
    to calculate the splitting scale
    in~\cite{Butterworth:2002tt,Butterworth:2007ke}, which takes the
    jet $\pt$ as the reference scale rather than the jet mass.}
\item Otherwise redefine $j$ to be equal to $j_1$ and go back to step 1.
\end{enumerate}
The final jet $j$ is to be considered as the candidate Higgs boson if
both $j_1$ and $j_2$ have $b$ tags. One can then identify $R_{b\bar
  b}$ with $\Delta R_{j_1j_2}$. The effective size of jet $j$ will
thus be just sufficient to contain the QCD radiation from the Higgs
decay, which, because of angular
ordering~\cite{Mueller:1981ex,Ermolaev:1981cm,Bassetto:1984ik}, will
almost entirely be emitted in the two angular cones of size $R_{b\bar
  b}$ around the $b$ quarks.

The two parameters $\mu$ and $y_\cut$ may be chosen independently of
the Higgs mass and $\pt$. Taking $\mu \gtrsim 1/\sqrt{3}$ ensures that if, in
its rest frame, the Higgs decays to a Mercedes $b\bar bg$
configuration, then it will still trigger the mass drop condition (we
actually take $\mu=0.67$).
The cut on $y \simeq \min(z_{j_1},z_{j_2})/\max(z_{j_1},z_{j_2})$
eliminates the asymmetric configurations that most commonly generate
significant jet masses in non-$b$ or single-$b$ jets, due to the soft
gluon divergence. It can be shown that the maximum $S/\sqrt{B}$ for a
Higgs boson compared to mistagged light jets is to be obtained with
$y_\cut \simeq 0.15$. Since we have mixed tagged and mistagged
backgrounds, we use a slightly smaller value, $y_\cut = 0.09$.


In practice the above procedure is not yet optimal for LHC at the
transverse momenta of interest, $\pt \sim 200-300\GeV$ because, from
eq.~(\ref{eq:Rbb}), $R_{b\bar b} \gtrsim 2\mh/\pt$ is still quite
large and the resulting Higgs mass peak is subject to significant
degradation from the underlying event (UE), which scales as $R_{b\bar
  b}^4$ \cite{Dasgupta:2007wa}.  A second novel element of our
analysis is to \emph{filter} the Higgs neighbourhood. This involves
resolving it on a finer angular scale, $R_\filt < R_{b\bar b}$, and
taking the three hardest objects (subjets) that appear --- thus one
captures the dominant $\order{\as}$ radiation from the Higgs decay, while
eliminating much of the UE contamination. We find $R_{\filt} =
\min(0.3, R_{b\bar b}/2)$ to be rather effective. We also require the
two hardest of the subjets to have the $b$ tags.

\begin{table}
  \centering
  \begin{tabular}{|l|r|r|r|}\hline
    Jet definition & $\sigma_{S}/$fb & $\sigma_{B}$/fb &
    $S/\sqrt{B\!\cdot\mathrm{fb}}$ \\ \hline
    C/A, $R=1.2$, MD-F         & 0.57 & 0.51 & 0.80 \\
    \kt, $R=1.0$, $y_{cut}$    & 0.19 & 0.74 & 0.22 \\
    SISCone, $R=0.8$           & 0.49 & 1.33 & 0.42 \\\hline
  \end{tabular}
  \caption{Cross section for signal and the $Z+$jets background in the
    leptonic $Z$ 
    channel for $200 < p_{TZ}/\!\GeV < 600$ and $110 < m_J/\!\GeV <
    125$, with perfect $b$-tagging; shown for our jet definition, and
    other standard ones at near optimal $R$ values.\vspace{-1em}}
  \label{tab:jet-perf}
\end{table}

The overall procedure is sketched in
Fig.~\ref{fig:higgs-neighbour+filter}. We illustrate its effectiveness
by showing in table~\ref{tab:jet-perf} (a) the cross section for
identified Higgs decays in $HZ$ production, with $\mh = 115\GeV$ and a
reconstructed mass required to be in an moderately narrow (but
experimentally realistic) mass window, and (b) the cross section for
background $Zb \bar b$ events in the same mass window. Our results
(C/A~MD-F) are compared to those for the \kt algorithm with the same
$y_\cut$ and the SISCone~\cite{Salam:2007xv} algorithm based just on
the jet mass. The \kt algorithm does well on background rejection, but
suffers in mass resolution, leading to a low signal; SISCone takes
in less UE so gives good resolution on the signal, however, because it
ignores the underlying substructure, fares poorly on background
rejection. C/A~MD-F performs well both on mass resolution and
background rejection.

The above results were obtained with
\herwig~6.510\cite{herwig1,herwig2} with {\sc Jimmy}~4.31~\cite{jimmy}
for the underyling event, which has been used throughout the
subsequent analysis. The signal reconstruction was also cross-checked
using~\pythia~6.403\cite{pythia}. In both cases the underlying event
model was chosen in line with the tunes currently used by ATLAS and
CMS (see for example~\cite{hlhc}~\footnote{The non-default parameter
setting are: %
PRSOF=0, JMRAD(73)=1.8, PTJIM=4.9~GeV, JMUEO=1, with CTEQ6L~\cite{Pumplin:2002vw} PDFs.
}).  The leading-logarithmic parton shower
approximation used in these programs have been shown to model jet
substructure well in a wide variety of
processes~\cite{Abazov:2001yp,Acosta:2005ix,Chekanov:2004kz,Abbiendi:2004pr,Abbiendi:2003cn,Buskulic:1995sw}. For
this analysis, signal samples of $WH, ZH$ were generated, as well as
$WW,ZW,ZZ,Z+{\rm jet},W+{\rm jet},t\bar{t}$, single top and dijets to
study backgrounds. All samples correspond to a luminosity $\ge
30$~fb$^{-1}$, except for the lowest $\ptmin$ dijet sample, where the
cross section makes this impractical. In this case an assumption was
made that the selection efficiency of a leptonically-decaying boson
factorises from the hadronic Higgs selection. This assumption was
tested and is a good approximation in the signal region of the mass
plot, though correlations are significant at lower masses.

The leading order (LO) estimates of the cross-section were checked by
comparing to next-to-leading order (NLO) results. High-$\pt$ $VH$ and
$Vb\bar b$ cross sections were obtained with
MCFM~\cite{Ellis:1998fv,Campbell:2003hd} and found to be about $1.5$
times the LO values for the two signal and the $Z^0b\bar b$ channels
(confirmed with MC@NLO v3.3 for the signal \cite{Frixione:2002ik}),
while the $W^{\pm}b\bar b$ channel has a K-factor closer to $2.5$
(as observed also at low-$\pt$ in
\cite{Campbell:2003hd}).\footnote{For the $Vb\bar b$ backgrounds these
  results hold as long as both the vector boson and $b \bar b$ jet have a high
  $\pt$; relaxing the requirement on $p_{TV}$ leads to enhanced $K$-factors from
  electroweak double-logarithms.}  The main other background, $t\bar
t$ production, has a K-factor of about $2$ (found comparing the \herwig\
total cross section to \cite{Nadolsky:2008zw}). This suggests that our
final LO-based signal/$\sqrt{\mathrm{background}}$ estimates ought not
to be too strongly affected by higher order corrections, though
further detailed NLO studies would be of value.

Let us now turn to the details of the event selection.
The candidate Higgs jet should have a $\pt$ greater than some
$\ptmin$.
The jet $R$-parameter values commonly used by the experiments are
typically in the range 0.4 - 0.7. Increasing the $R$-parameter
increases the fraction of contained Higgs decays.
%
%
Scanning the region
$0.6 < R < 1.6$ for various values of $\ptmin$ indicates an optimum
value around $R=1.2$ with $\ptmin = 200$~GeV. 

Three subselections are used for vector bosons: (a) An $e^+e^-$ or
$\mu^+\mu^-$ pair with an invariant mass $80 \GeV < m < 100\GeV$ and
$\pt > \ptmin$.  (b) Missing transverse momentum $> \ptmin$.  (c)
Missing transverse momentum $>30$~GeV plus a lepton ($e$ or $\mu$)
with $\pt > 30$~GeV, consistent with a $W$ of nominal mass with
$\pt > \ptmin$.  It may also be possible, by using similar techniques
to reconstruct hadronically decaying bosons, to recover signal from
these events. This is a topic left for future study.

To reject backgrounds we require that there be no leptons with $|\eta|
< 2.5, \pt > 30$~GeV apart from those used to reconstruct the leptonic
vector boson, and no $b$-tagged jets in the range $|\eta|<2.5, \pt >
50$~GeV apart from the Higgs candidate. For channel (c), where
the $t\bar{t}$ background is particularly severe, we require that
there are no additional jets with $|\eta| < 3, \pt > 30$~GeV. The
rejection might be improved if this cut were replaced by a specific
top veto~\cite{Butterworth:2002tt}. However, without applying the
subjet mass reconstruction to all jets, the mass resolution for $R=1.2$
is inadequate.

\begin{figure}[t]
  \begin{center}
  \includegraphics[width=1.0\linewidth]{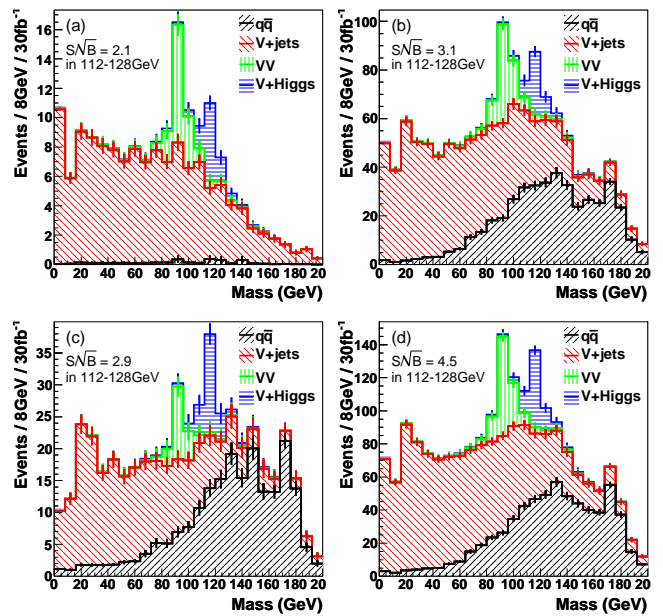}
  \caption{Signal and background for a 115~GeV SM Higgs simulated
  using \herwig, C/A~MD-F with $R= 1.2$ and $\pt > 200$~GeV, for 30
  fb$^{-1}$. The $b$ tag efficiency is assumed to be 60\% and a mistag
  probability of 2\% is used. The $q\bar{q}$ sample includes dijets
  and $t\bar{t}$. The vector boson selections for (a), (b) and (c) are
  described in the text, and (d) shows the sum of all three
  channels. The errors reflect the statistical uncertainty on the
  simulated samples, and correspond to integrated luminosities $>
  30$~fb$^{-1}$.\vspace{-1.5em}}
  \label{fig:optlep}
  \end{center}
\end{figure}

The results for $R=1.2, \ptmin = 200$~GeV are shown in
Fig.~\ref{fig:optlep}, for $m_H = 115$~GeV. The $Z$ peak from $ZZ$ and
$WZ$ events is clearly visible in the background, providing a
critical calibration tool. Relaxing the $b$-tagging selection would
provide greater statistics for this calibration, and would also make
the $W$ peak visible. The major backgrounds are from $W$ or $Z$+jets,
and (except for the $HZ (Z \rightarrow l^+l^-)$ case), $t\bar{t}$.

Combining the three sub-channels in Fig.~\ref{fig:optlep}d, and
summing signal and background over the two bins in the range
112-128~GeV, the Higgs is seen with a significance of $4.5~\sigma$
($8.2~\sigma$ for 100 fb$^{-1}$). The intrinsic resolution of the jet
mass at the particle level would allow finer binning and greater
significance. However, studies~\cite{sarah,sstef} using parameterised
simulations of the ATLAS detector indicate that detector resolution
would prohibit this.

\begin{figure}
  \begin{center}
  \includegraphics[width=1.0\linewidth]{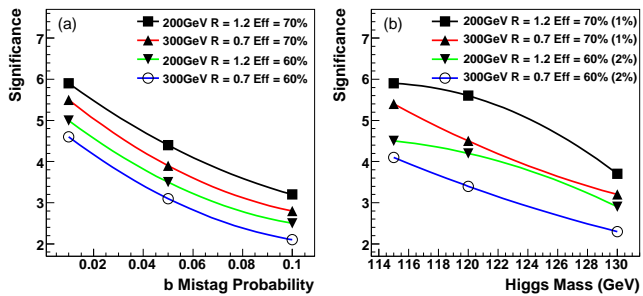}
  \caption{Estimated sensitivity for 30~fb$^{-1}$ under various
  different sets of cuts and assumptions (a) for $m_H = 115$~GeV as a
  function of the mistag probability for $b$-subjets and (b) as a
  function of Higgs mass for the b-tag efficiency (mistag rates) shown
  in the legend. Significance is estimated as ${\rm signal}/\sqrt{\rm
  background}$ in the peak region.\vspace{-1em}}
  \label{fig:curves}
  \end{center}
\end{figure}

The $b$-tagging and mistag probabilities are critical parameters for
this analysis, and no detailed study has been published of tagging two
high-$\pt$ $b$ subjets. Values used by experiments for single-tag
probabilities range up to 70\% for the efficiency and down to 1\% for
mistags. Results for 70\% and 60\% efficiency are summarised in
Fig.~\ref{fig:curves}a as a function of the mistag probability.

There is a trade-off between rising cross-section and falling fraction
of contained decays (as well as rising backgrounds) as $\ptmin$ is
reduced. As an example of the dependence on this trade-off, we show
the sensitivity for $\ptmin = 300$~GeV, $R = 0.7$ in
Fig.\ref{fig:curves}a.

The significance falls for higher Higgs masses, as shown in
Fig.~\ref{fig:curves}b, but values of $3 \sigma$ or above seem
achievable up to $m_H = 130$~GeV.

In addition to the $b$-tagging, the effects of pile-up, intrinsic
resolution and granularity of the detector will 
all have an impact. Several ideas exist to improve some of these, and
initial studies with realistic detector simulations indicate that the
efficiencies and resolutions assumed here are not unreasonable, though the exact
requirements of our analysis have not been studied with such tools.

We conclude that subjet techniques have the potential to transform the
high-$\pt$ $WH, ZH (H \rightarrow b\bar{b})$ channel into one of the
best channels for discovery of a low mass Standard Model Higgs at the
LHC. This channel could also provide unique information on the
coupling of the Higgs boson separately to $W$ and $Z$
bosons. Realising this potential is a challenge that merits
further experimental study and complementary theoretical investigations.

This work was supported by the UK STFC, and by grant
ANR-05-JCJC-0046-01 from the French Agence Nationale de la Recherche.
GPS thanks Matteo Cacciari for collaboration in extending FastJet to
provide the features used in this analysis.


\bibliographystyle{h-physrev4}

\bibliography{eurostar}

\end{document}